# Predictive Simulation: Using Regression and Artificial Neural Networks to Negate Latency in Networked Interactive Virtual Reality

Gregory Gutmann*, Akihiko Konagaya*

**Abstract**

Current virtual reality systems are typically limited by performance/cost, usability (size), or a combination of both. By using a networked client/server environment, we have solved these limitations for the client. However, in doing so we have introduced a new problem, namely increased latency. Interactive networked virtual environments such as games and simulations have existed for nearly as long as the Internet and have consistently faced latency issues. We propose a solution for negating the effects of latency for interactive networked virtual environments with lightweight clients, with respect to the server being used. The proposed method extrapolates future client states to be incorporated in the server's updates, which helps to synchronize actions on the client-side and the results coming from the server. We refer to this approach as *predictive simulation*. In addition to describing our method, in this paper, we look at extrapolation methods because the success of our predictive simulation method is dependent on strong predictions. We focus on regression methods and briefly examine the use of artificial neural networks.

## 1. Introduction and motivation

The motivation for this work is to negate the latency caused by our networked simulation environment in order to create a comfortable virtual reality (VR) experience. This is critical for VR because if the experience does not feel natural, such as when there are delays in movement or when motion continues after stopping, the user will likely feel uncomfortable, disoriented, or nauseous and not want to continue using VR.

If the situation was simply VR viewing of a networked simulation, we could use a streaming model. However, because we include interactions, latency becomes an issue. Furthermore, because the latency requirements of VR are very strict, our first effort is to minimize latency in a local area network (LAN) environment and consider the challenges that are involved. The acceptable latencies according to the type of game are listed below and shows how strict VR systems are in comparison to other games.

Acceptable Amounts of Perceivable Latency According to Game Type [1]:

- VR < 20 ms
- Monitor-based shooter games < 150 ms
- Real-time strategy games < 500 ms

The effects of latency are noticeable when (i) a client is interacting with server-side objects or (ii) two clients connected to the same server are interacting. In our first effort, although we do not explore multi-user latency issues, we keep them in mind when looking at the issues and potential solutions. In this work, we specifically examine the use of motion prediction to minimize the effects of latency in a client/server interactive VR environment where the client's physical actions are transmitted to the virtual environment.

This remainder of this paper is organized as follows: Section 2, the causes of latency; Section 3, background and previous work; Section 4, our predictive simulation approach; Section 5, testing methods; Section 6, the performance of various regression models for creating predictions; Section 7, the feasibility of using recurrent neural networks (RNNs) as predictors for our approach; Section 8, discussion; Section 9, future work; and Section 10, the conclusion.

## 2. Causes of latency and how we measure it

Latency is defined as "the amount of time a message takes to traverse a system" [2]. However, often it is used to describe round-trip time as well. This could be the ping time (i.e., the round-trip time for a small packet). Or in a networked virtual environment, it could be the time between a user command (input) and the result of that command being presented on the screen (output). Herein, we use the latter definition.

The general causes of latency are listed below along with the primary causes of latency within our own work.

---

*School of Computing, Tokyo Institute of Technology. 4259, Nagatsuta, Midori, Yokohama 226-8502, Japan

**General factors**

*Network-based*

- Processing delay of packets
- Transmission delay (bandwidth)
- Queuing delay
- Propagation delay (0.3 ns/m)

*Not network-based*

- Input sampling latency
- Render pipeline latency
- Simulation scale
- Display latency / pixel

**Primary factors that impact this work**

- Bandwidth
  - Initial baseline, then linear
- Network (distance, network device overhead)
  - Minimal on LAN
- Simulation scale
  - Initial baseline, then linear
- Input sampling latency (~10 ms)
  - Random impact between 0 and 10 ms
- Render pipeline latency (~11 ms)
  - Random impact between 0 and 11 ms

Looking at the list entitled "Primary factors that impact this work," the input sampling latency and render pipeline latency are random within a range that is based on (i) when the data are produced from the Leap Motion device with respect to the client application or (ii) when the data for rendering are prepared with respect to the state of the rendering pipeline. This random timing is due to the following three operations running asynchronously: input polling, networking, and rendering. The two aforementioned factors contribute to a latency floor, namely the lowest achievable latency.

As shown in Figure 1, the simulation scale and the transmission size, (i.e., the bandwidth), have a linear impact on latency, aside from a plateau at lower test cases. The plateau seen in the simulation scale (Figure 1a) is caused by the latency floor. The plateau seen in the bandwidth (Figure 1b) is due partly to the latency floor but also to the scaling behavior of network cards with respect to the transmission size until the card becomes saturated.

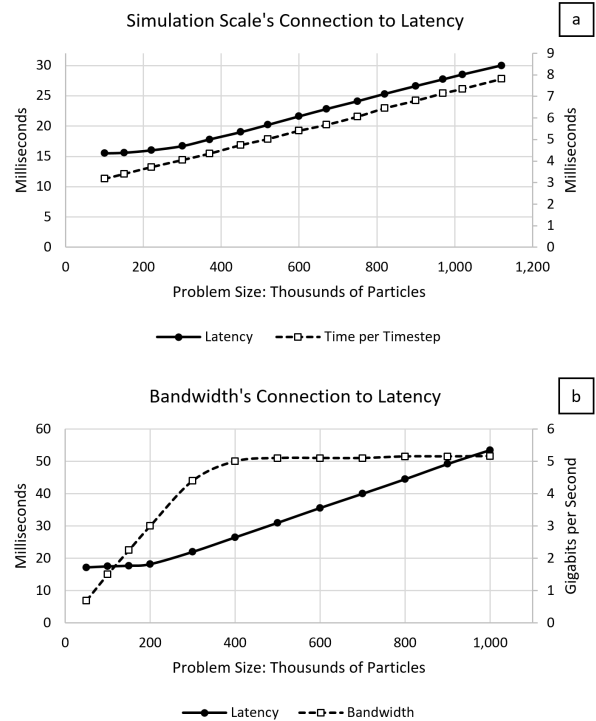

Figure 1. Impact of (a) simulation scale and (b) bandwidth on latency.

## 3. Background and previous work

Networked virtual environments have been around in various forms for quite some time. For example, one of the first networked games was Mazewar in 1973 [3]. Now there are countless networked virtual environments, including games, military simulations, and virtual collaboration environments. However, throughout the evolution of these technologies, latency has remained a persistent problem.

One approach that does not use any steps to negate latency is a client-server system in which the client is referred to as a **dumb client** or a dumb terminal [1]. This is when the client is generally treated as an input device and the server runs the simulation, which may be a game engine, a real-world environment, or a scientific simulation.

Many approaches have been taken to help negate the effects of latency. The following are a few examples.

With **client-side predictions**, the client also understands the dynamics of the virtual environment and can act (i.e., simulate) independently of the server [1]. This is commonly applied to games in which the dynamics being simulated are character motion and action in a mostly static environment. However, although this is a useful technique in some situations, it would go against our goal of offloading the simulation to the server. Also, in a highly dynamic environment, one user can have a large impact on the environment, thereby increasing the likelihood of local and server simulations diverging.

**Interpolation and local perception filtering** smooth out incoming state updates by interpolating across the received states instead of presenting them immediately to the client [4]. However, although this technique helps to create smooth visuals and compensate for discontinuities, it also masks an amount of latency that is equivalent to the age of the data being presented. Delaying results narrows the gap between action and result, but there is a limit to how far back in time the user can be put before creating discontinuity between user action and result.

In our work, although the user could be interpolated, interpolating all visible simulation data would place a moderate memory and computational load on the client. In our simulation, a user may be interacting with tens of thousands of particles, which may have impacts on hundreds of thousands of particles. Also, because we are working with a VR system, we are trying to avoid solutions that add latency.

**Lag compensation or timewarp** is a system that keeps a history of all recent player and environment states for a defined time interval, for example, 1 s in CS Source [5]. Then, if the user executes an action, the server looks back in time to the state of the world at the exact instant at which the user performed the action and simulates the action for verification [6,7]. However, although this is a very strong method for handling instant actions at a distance (e.g., shooter games), as the simulation size increases it becomes increasingly difficult to maintain a history of the simulation and step back through time. Also, because the actions in our simulation are/ primarily continuous ones (i.e., movement) and not instantaneous ones, there is currently no need for the capability that this technique provides.

**Extrapolation** takes older states and creates a predicted current or future state. If the user's actions can be predicted accurately enough, then the client could be a dumb client (i.e., an input device and a viewer), thereby keeping the simulation work on the server. Whereas interpolation is conservative, extrapolation is an optimistic algorithm, hoping to make correct predictions.

**Dead reckoning** is the process of predicting an entity's behavior based on the assumption that it will keep doing whatever it's currently doing [1,8]. This means extrapolation, and therefore discrepancies are almost certain and have been examined previously [9,10]. Nevertheless, there are benefits. Dead reckoning tries to minimize network bandwidth by sending state updates to the server only when the client deviates from the course that was last sent to the server. This is done in part by the client running a local simulation as well, which, as mentioned previously, is a problem for our intended use. When a client detects discrepancies, it can also interpolate from the current state to the correct state using an interpolation method that best fits the application, thereby maintaining a smooth experience. If the prediction error can be kept low, then this approach will help to keep the client(s) and server synchronized, owing to the predominant use of extrapolation instead of interpolation.

### 4. Our solution

Our chosen implementation, called **predictive simulation**, is based on extrapolation, where

1. client positions are sent to the server,
2. the server records a time series of position data,
3. the sever makes an extrapolated prediction into the future based on the client's round-trip latency,
4. the sever incorporates the results of the prediction into the simulation,
5. the server's results are sent to the client, and
6. the client renders the results, optionally overwriting the user particles received from the server with a local copy.

In this process, the server is authoritative and is solely responsible for manipulating the state of the virtual environment, thereby keeping all the simulation work on the server. The clients could be referred to as dumb clients because they do not simulate anything in their environment. However, as seen above, the clients can optionally override the simulation's results for their own input objects, such as particles making up their virtual hands, with a live version that is more up to date. This helps to remove some visual noise, thereby creating a more comfortable experience for the user in VR. For our work, this noise associated with the user's objects has a minimal impact on the environment, but this may not be the case for all simulations.

The clients could also optionally be put in charge of predicting their own future states rather than the server. This would remove the possibility of losing user position samples due to lost packets, but in-game environments may give the clients more opportunities to manipulate the system. As a side note, the steps listed above are simplified sequential representations of the steps actually taken; in the actual implementation, client and server act asynchronously.

If this approach were to be used without prediction, then the latency would be the time taken to run through steps 1 to 6, including some of the sources of latency listed in Section 2. To negate the latency with this approach, the server must act on future events (i.e., predictions) to keep the simulation results synchronized with the user's actions. Because of this, our primary challenge and focus is to reduce the perdition error as much as possible. In the following sections, we examine various extrapolation methods with the aim of minimizing the discrepancies between predicted future user states and the actual state.

*4.1. Predictors*

In practice, any predictor could be used with this approach. For example, one could choose interpolation such as linear interpolation or Lagrange interpolation (clarification: interpolation methods are often used for extrapolation), or one could use regression methods, a less common approach for games. Alternatively, from looking at recent trends, neural networks may be strong candidates and are introduced in Section 7.

Interpolation versus regression

- ***Interpolation*** *is the process of finding some predefined form that has the values of the n points provided exactly as specified.*
- ***Regression*** *is the process of looking for a function to fit a set of* n *points that minimizes some cost, usually the sum of the squares of the errors. The resulting function is not required to contain the exact values of the data but rather is an approximation of them.*

For our primary implementation, we have chosen to use linear and polynomial regression methods. We have done this because we consider each client position to be an estimate of the true position because of sampling noise (location tracking) and timing noise (fluctuating latency), which is in line with regression.

### 5. Methods

We chose to use a Leap Motion device for input because it has proven to be a very intuitive input device for VR demonstrations. Leap Motion is a dual-camera device with an application programming interface (API) that provides motion tracking of the palm and fingers. While this tracking technology is strong, a fair amount of noise due can be produced depending on environmental conditions (e.g., lighting, obstacles), occlusion (bending fingers with the palm facing the camera), the tracking algorithm, and the performance of the computer. This might seem counterintuitive given the goal of eliminating noise, but the strengths of the device justify its use.

Our process starts by recording 50 tracking points (x, y, z) for the two hands from Leap Motion's C API, which constitute the user input. To save network bandwidth, these points are then sent to the server to be expanded locally. The points are recorded on the server, and the matrix form $\widehat{\vec{\beta}} = (\mathbf{X}^\mathsf{T}\mathbf{X})^{-1}\mathbf{X}^\mathsf{T}\vec{y}$ of ordinary least-squares estimation is used for prediction [11]. This process is applied independently to the x, y, and z coordinates of all 50 points.

To form the hands in the simulation, the resulting predicted points are scaled and then filled in by interpolating across the points to generate hands with various numbers of particles. In our tests, one hand consisted of around 1300 particles. These data are then incorporated into the simulation and the results are sent to the user. During testing, we also generated a local particle hand from up-to-date Leap Motion data for comparison.

In the tests, we took steps to eliminate some noise and variance by removing the human element and automating the tests. For testing, we set up a fake hand that was held above the Leap Motion sensor. Because the Leap Motion device is usually attached to a head-mounted display (HMD), its motion is linked to that of the user's head. Therefore, we can move the camera in the virtual environment to automate moving the hand.

### 6. Performance

In this section, the error is defined as the instantaneous difference between two measured points, that is, the distance between them at any point in time. The two points being referred to are one a hand particle returned from the simulation, whose location may be the result of a prediction, and two a hand particle generated by the client using up-to-date input data.

*6.1. Tracking noise and reported error*

In addition to the factors outlined in Section 5 that contribute to tracking noise, the hands being generated on separate systems and with different client input samples across time add additional sources of noise when trying to compute the error rate. This results in having a non-zero minimum of for the error rate, due to a kind of noise floor.

The noise floor is evident in Figure 2a in the portions of the graph where the hand was motionless and no predictor was used. Under perfect conditions, the error rate should be zero when the hand is motionless because there are no changes with time.

Figure 2a and b showcase the potential variance in error based on factors such as the ones mentioned at the beginning of Section 5. Under non-ideal conditions, both the error rate and standard deviation increase, as is evident from Table 1.

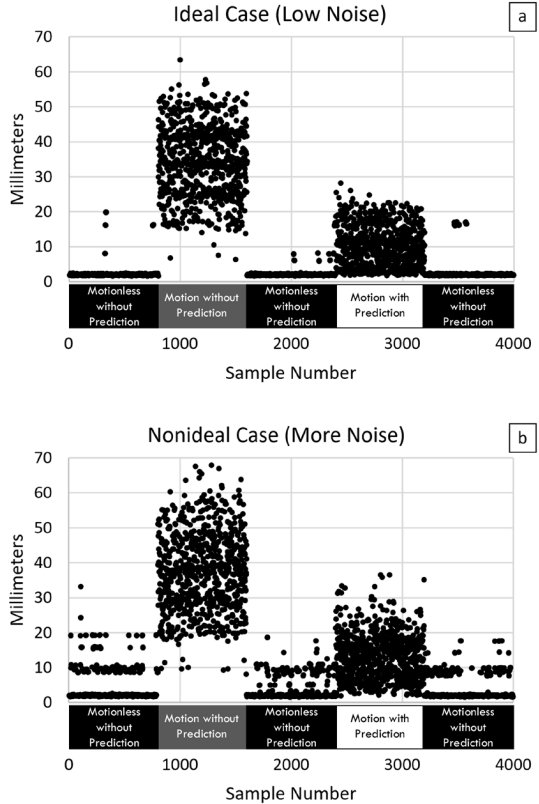

Figure 2. Tests of the error reported when motionless and when moving at a constant rate (recording was paused for a short time between each change in the testing situation). The y-axis is the average error per particle. When the prediction algorithm was enabled, linear regression with 10 data samples was used.

| Test | Condition | Average Error | Standard Deviation |
|---|---|---|---|
| Motionless with no prediction | Ideal | 2.39 | 2.13 |
|  | Nonideal | 4.21 | 3.94 |
| Motion with no prediction | Ideal | 33.87 | 10.15 |
|  | Nonideal | 37.74 | 11.40 |
| Motion with prediction | Ideal | 10.24 | 5.68 |
|  | Nonideal | 12.87 | 6.49 |

Table 1. Average errors and standard deviations for tests shown in Figure 2a and b.

*6.2. Testing regression perdition*

For testing, we wanted to test the robustness of the predictor with both constant motion and changes in motion, but we also wanted to test in a controlled and repeatable manner. Consequently, we decided on the back-and-forth motion outlined below with quick changes in direction. During normal use of our simulation, it is unlikely that the user will be moving at the test speed used and changing direction erratically, but it remains a possibility.

For our tests, we used the following:

- A single graphics processing unit (GPU) on the computer running as a server (any GPU count could be used if available)
- A single GPU on the computer running as a client for rendering
- A 10-gigabit LAN environment (never saturated due to focusing on low round-trip latency)
- Challenging prediction scenario:
  - 1 m/s back-and-forth hand motion
  - Changes in direction took place over 70 frames or around 0.53 s at 133 frames per second
  - Seven changes in direction in around 27 s
  - Changes in direction, over 0.53 s, were based on a sine function
    - *Timing of change in direction starts from initial deceleration from 1 m/s, including changing direction, then stops when reaching 1 m/s again in the opposite direction*
- Leap Motion as the input device
  - New samples being created every 11 ms
- Polynomial regression for extrapolation (predictions)
  - First-order (linear) with various numbers of sample points
  - Second-order with various numbers of sample points
  - Third-order was tested but was worse than having no predictor, so it was omitted

The results of the test are shown in Figure 3. Linear regression is stable at lower sample counts, but the error rate increases as more samples are used. As more samples are used with a linear predictor, the predictor's ability to handle rapid changes in movement decreases. Second-order polynomial regression is unstable at low sample counts because of the noisy nature of the raw data. It is attempting to predict the noise, but as the sample count increases the predictor is better able to predict the true motion path. In this test, the second-order polynomial regression is shown to be stronger over a wide range, namely 10 to 40 samples.

In practice, if it is known that the user will constantly make rapid changes in direction, then we suggest using lower sample counts in that range; with less memory of past states, the regression adapts more readily to changes. If the data are noisy, then we suggest using higher sample counts within that range; with more memory of past states, regression has a normalizing effect.

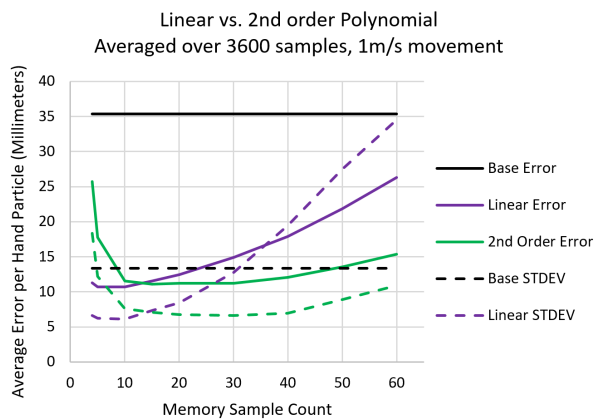

Figure 3. Solid lines correspond to average error per particle in the hand, a measure of the predicted hand's error with regard to position and orientation. Dotted lines correspond to the standard deviation of the error for the given test. Memory samples refer to the number of location samples over time that were used for the regression-based prediction. The base case (i.e., no prediction) is shown in black. Linear prediction is shown in purple, and second-order polynomial prediction is shown in green. The individual tests that this graph is based on can be found in the appendix.

Based on the tests in Figure 3, the best sample count for linear regression is seven samples, which reduced the error by a factor of 3.31 and the standard deviation by a factor of 2.15 with respect to the base case. While this will work in a LAN environment, seven samples here is a sampling of only 70 ms of user action, which would likely be too short to compensate for lag spikes or lost packets in a wide-area network (WAN) environment. Therefore, we chose second-order regression with 20 samples as the best candidate for our work, which reduced the error by a factor of 3.16 and the standard deviation by a factor of 2.00. A visual example is shown in Figure 4.

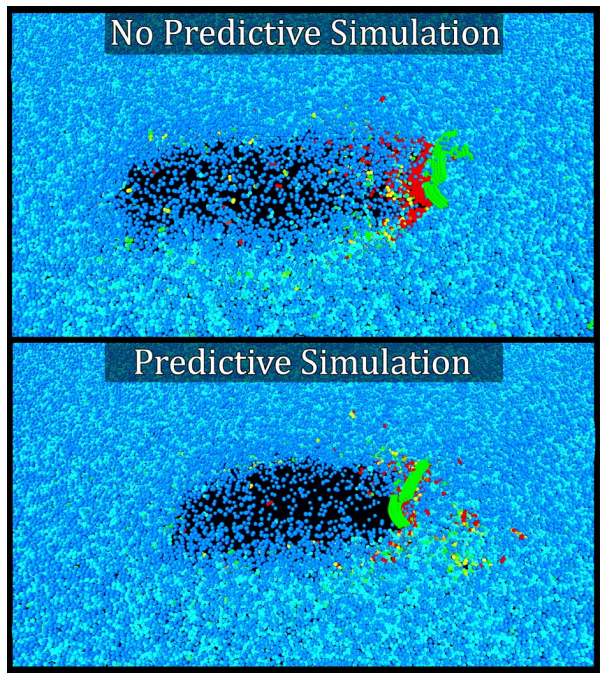

Figure 4. In the image with no prediction (top), the strongly interacting particles in red can be seen behind the hand. In the image using our predictive simulation method (bottom), the strongly interacting particles are cupped in the hand. The amount of impact our method has is hard to portray outside of VR, see page one a link to videos for a better representation.

## 7. Ongoing work: neural networks for prediction

With the recent excitement around artificial intelligence (AI) and specifically artificial neural networks (ANNs) to solve problems, many people have explored the use of ANNs and have been successful in their efforts. Interesting and strong work on human motion prediction is that by Martinez et al. [12]. However, they looked at predicting defined human actions, whereas we are trying to predict motion where there is no defined action.

The simulated objects in our work often vary greatly in shape, size, and dynamics. As such, we took a different route and made an initial attempt at using simpler RNNs, testing gated recurrent unit (GRU) and long short-term memory (LSTM) networks.

Our starting point was experimenting with RNNs in Keras for simplicity and using code from Aungiers' GitHub as a basis [13]. After some modification, we could use the code to generate predictions of user movement several data points (i.e., time steps) in the future. We then converted the trained model to be run by TensorFlow's C API within our code. However, although this approach works, it is a bit disconnected with respect to training loss and the error seen in VR caused by latency. Training the network live to remove this disconnect remains as future work.

Our initial ANN approach was to train separate models for each varying amount of latency, namely a model for 10 ms, 20 ms, 30 ms, and so on. This simplifies the model to create our initial proof of concept more easily, but it is far from ideal. Our future efforts will involve training one model that is aware of the latency (i.e., the desired prediction distance) so that the model will work efficiently at any latency value or with fluctuating latency.

The results of our initial work with RNN prediction are shown in Figure 6, which shows a test containing around 40 ms of latency. For the following tests, we used a different measurement of error than that in Section 6 because particle indexing differed dramatically between live data and prediction when using the ANN. The measurement used here is a summation of the minimum particle distances between each predicted point to any of the points in the live data. Although it is not a perfect measurement (e.g., offset fingers could overlap if the error is high), it gives a sense of the separation between the predicted hand and the live hand. This measurement is used only to compare the prediction methods, not to assess the reduction in error rate. Figure 5 shows what the error rate looks like with no prediction used as a reference.

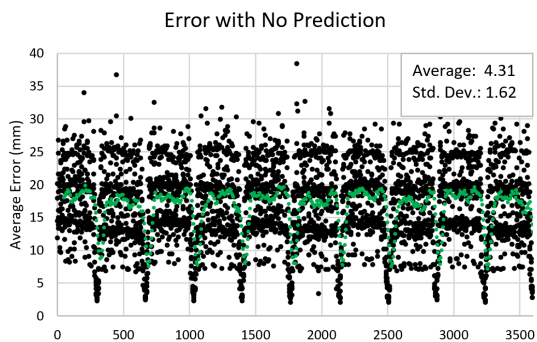

Figure 5. The reported error rates are lower than in Figure 3 due to overlapping adjacent fingers. The x-axis is the sample number.

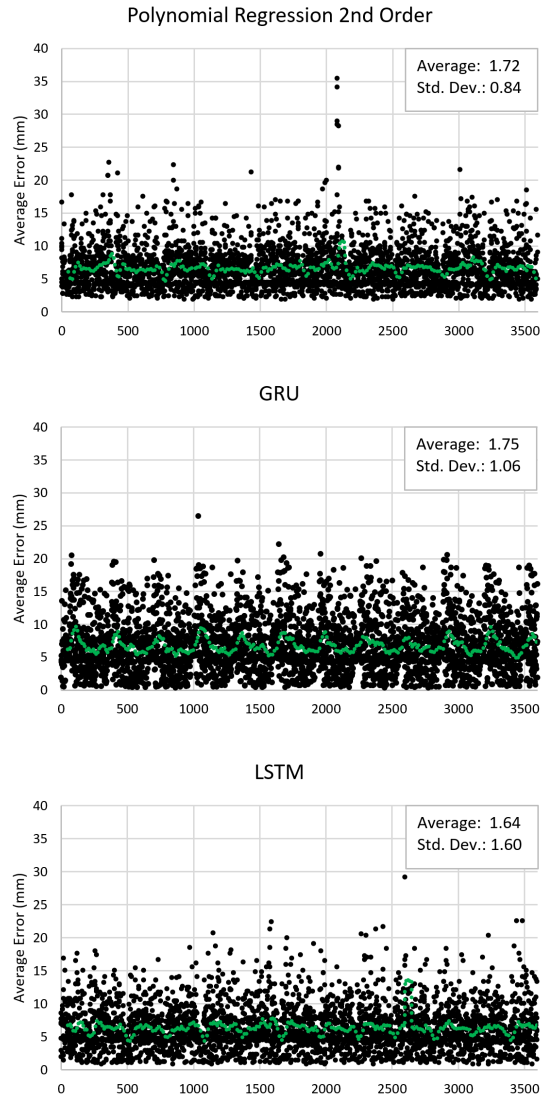

Figure 5. Differences between prediction methods. The x-axis is the sample number. The green line shows a moving average of 50 samples. The GRU and LSTM networks both had three types of layer, namely 60 input layers, 10 hidden layers, and one output layer. The training data consisted of around 20,000 samples.

With no prediction, the back-and-forth movement can be seen, with drops in error as the hand slows to change direction. When using GRU, the pattern is inverted, with a lower error during constant movement and higher error during changes in direction, similar to that with a linear predictor. More effort in tuning the network should reduce the slightly linear behavior. Lastly, the polynomial regression and LSTM performed fairly similarly, although the LSTM had a slightly lower average error and the regression had a lower standard deviation.

In comparison with our regression implementation, our simple RNN required extensive development time. However, in contrast to cutting-edge ANNs in other research fields, our current RNN is just scratching the surface of the potential capabilities of ANNs.

## 8. Discussion

In Table 2, we have outlined the key attributes of our method and implementation of predictive simulation for reducing the effect of latency.

| Attributes | Further Details |
|---|---|
| Offloads simulation/environment to the server | Doesn't require a local simulation to negate latency |
| | Can handle large dynamic environments with light clients |
| Brings user as close as possible to live interaction with the server | Using extrapolation instead of interpolation |
| Computes environment state updates based on predictions | Objects in the environment react to the user's future action; therefore, the resulting environment state becomes a representation of a future state. However, the state doesn't reach the client until the environment state time matches the live client time, due to latency. |
| Server is in full control of the environment: security | Client does not directly make any changes to their environment since they are not given the capability of simulating the environment. |
| Does not offload the rendering to the server (if offloading due to scale) | Local rendering ensures near-zero latency with camera movement. Critical for VR comfort |
| | Highly dynamic scenes (particles) do not compress well, reducing the benefit of server-side rendering with high-resolution VR HMD |

Table 2. Attributes and details of the present work.

To generalize, our method is most beneficial when applied to an interactive environment for which the computation must be offloaded. Reasons for offloading may include performance or security. Predictive simulation is needed in those cases because if all computational work is offloaded, then the client cannot locally take steps to negate the effects of latency, so steps must be taken on the server (simulation side). Predictive simulation can, therefore, be used to keep the client and server synchronized by removing discrepancies between client actions and the result of the actions from the server. It could also be used to achieve beneficial results in other situations that do not require full offloading. However, other methods such as dead reckoning would also be applicable in those cases.

More-specific examples include:

- Low-power clients networked to other systems (PC, server, cloud)
  - Client/server VR
  - Gaming as a service (GaaS)
- Networked simulation
  - Shared internal use in an organization
  - Cloud-based
  - Simulation as a service (SaaS)
- Networked games with highly dynamic environments

Thinking further beyond our own work, there is a potential for use with drones (land, air, sea, space). Drones often operate in high-latency environments that can be simulated using various scene-reconstruction methods that are currently being developed [14-16].

The proposed method has some limitations at present: the environment or focus of the work must be able to be simulated or tracked to some degree, the method does not work for actions with instant results, and it is not currently designed for synchronizing multiple users in competitive environments (i.e., games).

### 8.1. Predictors

Regarding the predictor that is used, both regression and AI methods have positives and negatives. Depending on the situation, one may be better than the other.

Regression

- Adaptive, every prediction is generated from very recent data
- Struggles to predict long distances
- Struggles to predict erratic movement
- Fast to compute (<1 to 2 ms)
- Short development time
- Analytic solution

ANN

- Can draw on past experience for predictions
- Can adapt across experiences but not always beyond them
- Can use more data types than user position and environment
- Can adapt to users or use cases
- Relatively fast (2 to 4 ms)
- Long development time
- Analogy: ideally, like human intuition

## 9. Future work

We plan to experiment with WANs and test the feasibility of these methods in a less-predictable network environment.

As our work with ANNs progresses, we plan to test more extensively, such as with multiple different users or using VIVE VR trackers with custom VR objects. Extensive testing was not as necessary with regression because it is always adapting to the situation with no memory (bias) regarding specific movements, as ANNs may have.

We also plan to extend our work to handle multiple users. With our approach, multiple users can be kept in synchronization with the environment naturally. However, creating consistency between users in the environment is a non-trivial issue because each user may have a different round-trip latency with the server, thereby creating unique latency values between each pair of users. A method similar to "timelines" may be a possible solution as long as client computation can be minimized when there are significant numbers of objects [17].

Lastly, minimizing sources of latency that are within our control is also a constant effort but is beyond the scope of work.

## 10. Conclusion

In this work, we have explored the latency issues in our networked VR environment and proposed a method that we call *predictive simulation* as a solution to negate the latency. It has been designed for use in interactive environments for which computational work is offloaded. With this approach, we can dramatically decrease the perceived latency in our simulation, reducing the difference between data that have gone to the server and back versus live client-side data. When using regression, the error rate has been reduced by a factor of more than three and the standard deviation by a factor of two, all while the client is interacting with tens to hundreds of thousands of dynamic objects on the server. We also introduced our initial work using ANNs as predictors, which we believe will have greater potential as predictors over regression in future efforts.

### Acknowledgements

This research was partially supported by the Strategic Advancement of Multi-Purpose Ultra-Human Robot and Artificial Intelligence Technologies of the New Energy and Industrial Technology Development Organization (NEDO, P15009) of Japan, MEXT/JSPS KAKENHI Grant Number 17H00769 and 18H03673.